\documentclass[fleqn,twoside]{article}
\usepackage{espcrc2}

\usepackage{graphicx}
\usepackage[figuresright]{rotating}

\newcommand{\AmS}{{\protect\the\textfont2
  A\kern-.1667em\lower.5ex\hbox{M}\kern-.125emS}}

\title{Strangelets in cosmic rays}

\author{M.Rybczy\'nski\address[a]{Institute of Physics,
        \'Swi\c{e}tokrzyska Academy, Kielce, Poland}
        \thanks{e-mail: mryb@pu.kielce.pl},
        Z.W\l odarczyk\addressmark[a]\thanks{e-mail: wlod@pu.kielce.pl}
        snd
        G.Wilk\address{The Andrzej Soltan Institute for
        Nuclear Studies, Nuclear Theory Department, Warsaw, Poland}
        \thanks{e-mail: wilk@fuw.edu.pl}}

\begin{document}

\begin{abstract}
Recently new data from the Cosmo-LEP project appeared, this time from
DELPHI detector. They essentially confirm the findings reported some
time ago by ALEPH, namely the appearence of bundles of muons with
unexpectedly high multiplicities, which so far cannot be accounted by
present day models. We argue, using arguments presented by us some
time ago, that this phenomenon could be regarded as one more
candidate for the presence in the flux of cosmic rays entering the
Earth's atmosphere from outer space nuggets od Strange Quark Matter
(SQM) in form of so called {\it strangelets}.
\vspace{1pc}
\end{abstract}

\maketitle

\section{INTRODUCTION}

Recently new data from Cosmo-LEP program, this time from DELPHI
detector, has been reported \cite{CosmoDelphi}. Among other things
they have confirmed the findings reported before by ALEPH
\cite{CosmoAleph}, namely that one observes bunches of cosmic muons
(i.e., produced at the top of the Earth's atmosphere) of unexpected
large multiplicities (up to $N_{\mu} = 150$). Their origin is so far
unexplained and no model used in Monte Carlo (MC) programs simulating
cascades of cosmic rays (CR) in the atmosphere is able to account for
this phenomenon. In \cite{CosmoDelphi} the expectation was made that
source of this discrepancy can eventually come directly from the
elementary interaction model used in MC. However, in our opinion,
which we would like elaborate here in more detail, it could rather
come (at least to a large extent) from the projectile initiating the
cascade. Namely, as we have already done in many places on other
occassions \cite{S,RWWAPPB}, we shall argue that the abovementioned
results of both experiments can be regarded as yet another signal of
the presence in the flux of CR entering the Earth's atmosphere of
nuggets of Strange Quark Matter (SQM) called {\it strangelets}. In
this way results of \cite{CosmoDelphi} and \cite{CosmoAleph} would
just continue a long list of other phenomena explanable in this way
like anomalous cosmic ray burst from {\it Cygnus X-3}, extraordinary
high  luminosity gamma-ray bursts from the {\it supernova remnant
N49} in the Large Magellanic Cloud or {\it Centauro} (to mention only
the most interesing and intriguing examples, for more details see
\cite{RWWAPPB,SS} and references therein). In \cite{RWW} we have
already provided successful explanation of ALEPH observations by
using notion of strangelets and assuming their flux being the same as
obtained from analysis of all previous signals of strangelets present
in the literature. (Actually, at that time ALEPH results were
circulated only as conference papers, however, the final results
presented in \cite{CosmoAleph} turned out to be identical to those
addressed in \cite{RWW}). 

It is worth to remind here that CosmoLEP data are very important
because: $(a)$ the high multiplicity cosmic muon events ({\it muon
bundles}) are potentially very important source of information about
the composition of primary CR because muons transport in
essentially undisturbed way information on the first interaction of 
the cosmic ray particle with atmosphere; $(b)$ such events have never
been studied with such precise detectors as provided by LEP program
at CERN, nor have they been studies at such depth as at CERN
\cite{LEP} (ranging between $30$ and $140$ meters what corresponds to
muon momentum cut-off between $15$ and $70$ GeV).

\section{SOME FEATURES OF STRAN\-GE\-LETS}

For completeness let us remind here the most important for us
features of strangelets (see \cite{S,RWWAPPB} for details). They are
hadron-like being a bag of up, down and strange quarks (essentially
in equal proportion) becoming absolutely stable at high mass number
$A$ (more stable than the most tightly bound nucleus as iron).
However,  they become unstable below some critical mass number,
$A_{crit} = 300 - 400$. Despite the fact that their geometrical radii
are comparable to those of ordinary nuclei of the corresponding mass
number $A$, $R=r_0 A^{1/3}$, they can still propagate very deep into
atmosphere. This is because \cite{S} after each collision with the
atmosphere nucleus strangelet of mass number $A_0$ becomes just a new
\vspace{-10mm}
\begin{figure}[h]
\setlength{\unitlength}{1cm}
\begin{picture}(8.0,8.0)
\includegraphics{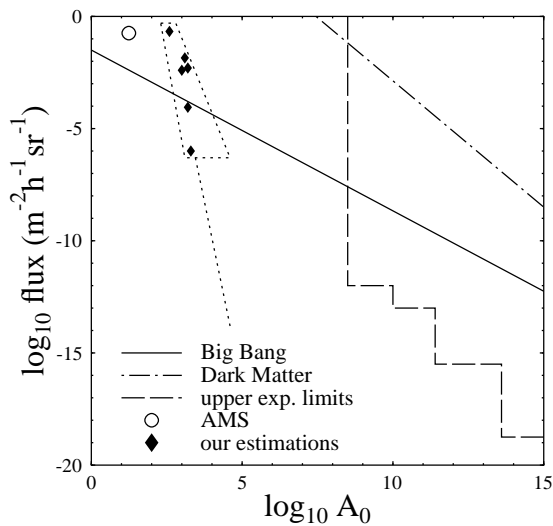} 
\end{picture}
\vspace{-15mm}
\caption{The estimated flux of strangelets \protect\cite{RWWAPPB} 
compared with existing upper experimental limits \protect\cite{Price}
and with other predicted astrophysical limits.}
\end{figure}
strangelet with mass number approximately equal $A_0-A_{air}$ and
this procedure continues unless either strangelet reaches Earth or
(most probably) disintegrates at some depth $h$ of atmosphere
reaching $A(h)=A_{crit}$. Actually, in a first approximation (in
which $A_{air} << A_{crit} < A_0$), in the total penetration depth of
the order of 
\begin{equation}
\Lambda\, \simeq\, \frac{4}{3}\, \lambda_{N-air}\,
\left(\frac{A_0}{A_{air}}\right)^{1/3} \label{eq:L}
\end{equation}
where $\lambda_{N-air}$ is the usual mean free path of the nucleon in
the atmosphere. 

There are number of candidates for strangelets known in the
literature, the common feature is their small ratio of charge $Z$ and
mass $A$ numbers, $Z/A$. The so called {\it Saito events} have
$Z\simeq 14$ and $A\simeq 350$ and $A\simeq 450$ \cite{Saito}. The
most spectacular is {\it Price event} \cite{Price} with $Z \simeq 46$
but $A > 1000$. On the other hand the {\it Exotic Track event} (ET)
\cite{ET} has been produced after the respective projectile has
traversed $\sim 200$ g/cm$^2$ of atmoshere. Finally, the so called
{\it Centauro events} \cite{Centauro} has been produced at depth 
$\sim 600$ g/cm$^2$ and contains probably $\sim 200$ baryons
\cite{MyC}. In Fig. 1 we show the resulting flux of strangelets
obtained by considering the above signals \cite{RWWAPPB}. One can add
to them the recently registered with AMS detector \cite{AMS} event
with small ratio $Z/A$ and also very small $A$, estimated to be
$A\simeq 17.5$, it could be a metastable strangelet.

\section{RESULTS}

This is the picture we shall use to estimate the production of muon
bundles produced as result of interaction of strangelets with
atmospheric nuclei. We use for this purpose the SHOWERSIM
\cite{SHOWER} modular software system specifically modified for our
present purpose. Monte Carlo program describes the interaction of the
primary particles at the top of atmosphere and follows the resulting
electromagnetic and hadronic cascades through the atmosphere down to
the observation level. Registered are muons with momenta exceeding
$70$ GeV for ALEPH and $50$ GeV for DELPHI. Primaries initiated
showers were sampled from the usual power spectrum $P(E) \propto
E^{-\gamma}$ with the slope index equal to $\gamma = 2.7$ and with
energies above $10\cdot A$ TeV.  

The integral multiplicity distribution of muons from ALEPH data are
compared with our simulations in Fig. 2. For completeness DELPHI data
are present also. At first we have used here the so called "normal"
chemical composition of primaries with $40$ \% of protons, $20$ \% of
helium, $20$ \% of C-N-O mixture, $10$ \% of Ne-S mixture and $10$ \%
of Fe. As one can see in Fig. 2 it can describe only the low
multiplicity ($N_{\mu} \le 20$) region of ALEPH data. The small
admixture of strangelets with the mass number $A=400$ being just
above the estimated critical one estimated $A_{crit}\sim 320$ in the
primary flux of CR (corresponding to relative flux of strangelets
$F_S/F_{total}\simeq 2.4\cdot 10^{-5}$) can, however, fully
accomodate ALEPH data. As can be noticed DELPHI data
\cite{CosmoDelphi} differ rather substantially in shape from ALEPH
data. They could be described equally well for $N_{\mu} > 40$ but
only with $5$-fold smaller flux of strangelets. However, in this case
events with small $N_{\mu}$ would fall completely outside the
fit\footnote{In fact nothing better can be done because neither ALEPH
or DELPHI can at the moment provide any explanation of this visible
discrepancy of their respective results.}. 
\vspace{-10mm}
\begin{figure}[h]
\setlength{\unitlength}{1cm}
\begin{picture}(8.0,9.5)
\includegraphics{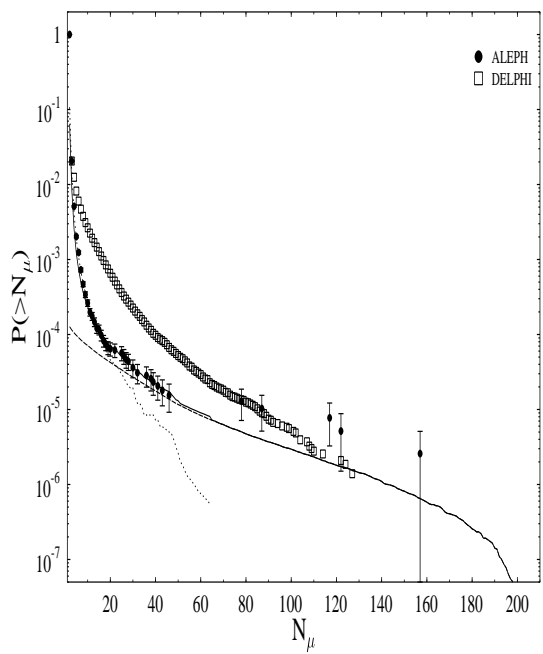} 
\end{picture}
\vspace{-15mm}
\caption{Our results (dotted line corresponds normal composition of compared with ALEPH data
\protect\cite{CosmoAleph}. Notice that there is big discrepancy
between ALEPH and DELPHI data \protect\cite{CosmoDelphi}, shown here
for comparison.}
\end{figure}
\vspace{-10mm}

One should notice that  results of both experiments differ already at
small values of muon multiplicity. It looks like DELPHI makes
preference for heavy composition of primary CR right from the
beginning whereas ALEPH preferes somohow lighter (protonic)
composition of CR. In any case, the excess of muons is clearly
visible therefore we regard this as a possible additional signal of
strangelets\footnote{There are still expected data from L3
experiment, however, so far the muonic part is not yet ready
\cite{L3}.}. 

\section{CONCLUSIONS}

To conclude: we propose to regard the Cosmo-LEP data on CR muons
obtained so far as an additional possible signal of the possible SQM
admixture present in the primary CR flux. We would like to add here
that such admixture would also contribute to CR flux at energies
greater than GZK cut-off \cite{RWWAPPB,M} explaining therefore this
phenomenon in a quite natural way\footnote{If it will be finally
confirmed by experiment \cite{GZK}.}. This makes strangelets
interesting subject to investigate in the future.

We would like to close with the following remark. With the flux of
strangelets as estimated by us and used here (equal to $F_S/F_{total}
= 2.4\cdot 10^{-5}$ in the enrgy range of tens of GeV) the energetic
spectrum of strangelets should fall like $\sim E^{-2.4}$, i.e., with
spectral index being much smaller than for protons. Actually, this
result agrees nicely with $A$-dependence of the spectral index of
CR's obtaine when fitting the world CR data \cite{SI}.

\end{document}